\documentclass[12pt]{iopart}

\begin{document}

\title[Plane wave evolution in a square barrier]{Infinite plane wave evolution in a 1-D square
quantum barrier}

\author{J. Julve}

\address{IMAFF, Consejo Superior de Investigaciones
Cient\'\i ficas, Serrano 113 bis, Madrid 28006, Spain}

\ead{julve@imaff.cfmac.csic.es}

\author{F. J. de Urr\'{\i}es}

\address{Departamento de F\'\i sica,
Universidad de Alcal\'a de Henares, Alcal\'a de Henares (Madrid),
Spain}

\ead{fernando.urries@uah.es}

\begin{abstract}
We analytically compute the time evolution of an initial infinite
plane wave in the presence of a 1-dimensional square quantum
barrier. This calculation generalizes the analysis of the shutter
problem and sets the basis for the calculation of the transmission
of general wave packets, aiming to work out the explicit
contribution of the resonant (Gamow) states. The method relies
mainly on the analytical properties of the Green function. The role
of separate boundary conditions on the Green function and on the
evolution equation is highlighted. As in previous works on related
problems, only the determination of the resonant momenta requires
numerical methods.


\submitto{Journal of Physics A: Mathematical and General}

\end{abstract}

\maketitle

\section{Introduction}

Seemingly simple quantum mechanical problems like the traversal of
potential barriers by wave packets are surprisingly rich of
mathematical, technical and physical details. The 1-dimensional case
is interesting in its own: electron transport through barrier
junctions, the physics of light and wave guides, particularly of the
transmission through Photonic Band Gaps, the time of arrival of wave
packets and the ensuing paradoxes of non-locality and super-luminal
tunnelling are examples.

Because of its strong overlapping with the standard theory of 3-D
scattering, usually the s-wave radial problem for the spherical
shell potential is studied in the literature as an example of the
role of the potential barriers in connection with the evolution of
decaying states. These states obey specific boundary conditions
(BCs), namely the vanishing of the wave function at the origin and a
purely outgoing BC outside the barrier. However the corresponding
BCs adopted in the 1-D scattering make it different also from a
mathematical point of view, providing a different setting for the
related resonant (Gamow) states.

On the other hand, practical calculations face important limitations
and the effort of pushing the analytical calculation as far as
possible is always rewarding. Significant progress has been
accomplished \cite{Muga1} for a Gaussian packet impinging on a
square barrier, where the Gaussian structure was exploited, and
\cite{Calderon1} for the shutter problem, where the contribution of
the resonant states in a double square barrier was worked out.
Relevant resonances are mostly expected in systems of two or more
barriers separated by a gap. Actually they generally occur in any
simple barrier, and the plain square barrier is most tractable,
fully representative for many theoretical purposes and devoid of
bound states or other unessential features.

In this paper we consider the time evolution, in the presence of a
square barrier, of an initial plane wave with support on the whole
1-D space, aiming to work out the contribution of the resonant
states. These states have a central role, as long as they correspond
to complex momenta (or energies) which are poles of the S-matrix,
and dictate the most convenient type of BCs to be used in the
problem.

Our calculation extends the results of shutter problem, where the
initial plane wave occupied the space axis at one side of the
barrier and was specially suited to study details of the propagation
of the wave front. We provide a comprehensive result containing also
the evolution of the segments of the initial plane wave lying inside
and at the other side of the barrier.

Our main goal however, is to pave the way for further uses since
superpositions of infinite plane waves build up any desired initial
wave packet. In particular, the compact result obtained for the
Gaussian wave packet \cite{Muga1} could be re-derived pinpointing
the contribution of each single Fourier component.

A byproduct of our calculation is to provide an explicit example
where we can study the properties of completeness of the resonant
states in different regions of the square potential, an issue
properly dealt with \cite{Madrid1} in the frame of Rigged Hilbert
Spaces, which will be addressed elsewhere.

In this paper we follow the approach used by Peierls and
Garc\'{\i}a-Calder{\'o}n \cite{Calderon2}, namely
Laplace-transforming the time evolution equation into a second order
linear differential equation (SOLDE) in one variable, expressing its
solutions in terms of the Green function (GF) with resonant boundary
conditions (RBCs) and then undoing the Laplace transform.

The first step is done in Section 2, where a brief discussion is
made of the use of the Green's method for the solution of SOLDEs
when the BCs required for the GF and for the solution are different.
This point is relevant for the issue of the completeness of the
Gamow states. The GF with RBCs features a simple structure of
isolated (resonance) poles, in terms of which the ($p$-dependent)
Laplace-transformed wave function is worked out in Section 3
previously to the second step. Then the explicit calculation of the
inverse Laplace transform leading to the sought after $t$-dependent
solution is carried out in Section 4.

As a check of this cumbersome analytical calculation, the $t=0$
(Section 5 of the paper) and the $t=\infty$ (Section 6) limits of
the time-dependent solution $\psi (x,t)$ are calculated: at $t=0$
one must recover the initial plane wave and at $t=\infty$ a suitable
stationary solution must be reached. Then the conclusions are drawn
in Section 7.

Some notations and a number of calculations and technical details
are deferred to the Appendices in an effort to make the paper more
self-contained.

\section{The evolution equations}

We consider the 1-D time-dependent $\rm Schr\ddot{o}dinger$ equation
\begin{equation}
(\rmi\hbar\frac{\partial}{\partial t}-H)\psi (x,t)=0
\end{equation}
where the Hamiltonian $H=p^2/\,2m\;+V(x)$ corresponds to the square
barrier potential $V(x)=\theta(x)\theta(L-x)V$ and the solution
satisfies the initial condition $\psi (x,0)\equiv\psi_o(x)=
\rme^{\rmi kx}\;,\,k>0\, $.

The Laplace transform $\bar{\psi}(s)=\int_o^\infty \rmd t\,\,
\rme^{-st}\psi(t)$ on the time variable, brings the parabolic
partial derivative differential equation (1) to the simpler SOLDE
\begin{equation}
[\frac{\partial^2}{\partial
x^2}+p^2-\frac{2m}{\hbar^2}V(x)]\;\bar\psi(x,p)
=\rmi\alpha\;\rme^{\rmi kx}
\end{equation}
 where $\alpha\equiv 2m/{\hbar}$ and $p^2=\rmi\alpha\,s$ (Appendix A).  The
homogeneous equation $L_x\,\bar\psi=0$, namely the time-independent
$\rm Schr\ddot{o}dinger$ equation for the potential $V(x)$, has a
variety of solutions according to the BCs we choose at $x=0$ and
$x=L$ depending on the physical problem. Besides the usual
scattering \emph{in} and \emph{out} solutions for continuous and
real $E=p^2/\,2m\,>0$, one has the resonant solutions satisfying the
homogeneous outgoing RBCs
\begin{equation}
\partial_x \bar\psi\!\!\mid_{x=0}\,=-\rmi p\,\bar\psi(0) \;\;,\;\;
\partial_x \bar\psi\!\!\mid_{x=L}\,=\rmi p\,\bar\psi(L)
\end{equation}
(the ingoing reversed sign of $p$ is not considered here for the
reasons outlined later on). These solutions exist only for a
denumerable set of isolated values $p_n$ of $p\;$ lying in the lower
half complex plane (Appendix A).

A solution to the inhomogeneous equation (2) for $p^2 \in R_{+}$ can
be explicitly written down for each region of $V(x)$ :

\begin{eqnarray}
\bar\psi_{I}(x,p)&=&B\,\rme^{-\rmi px}+\frac{\rmi\alpha}{p^2-k^2}\rme^{\rmi kx}\hskip 2.7cm(x<0)\nonumber\\
\bar\psi_{II}(x,p)&=&M\,\rme^{\rmi p'x}+N\,\rme^{-\rmi p'x}+\frac{\rmi\alpha}{p'^2-k^2}\rme^{\rmi kx}\hskip 0.5cm(0\leq x\leq L)\;\\
\bar\psi_{III}(x,p)&=&A\,\rme^{\rmi px}+\frac{\rmi\alpha}{p^2-k^2}\rme^{\rmi kx}\hskip 3cm(x>L)\nonumber\\
\nonumber
\end{eqnarray}
where the amplitudes $A$, $B$, $M$ and $N$ are functions of $p$
completely determined by the matching conditions at $x=0$ and $x=L$
and have a common denominator of the form
$D(p)\equiv\ominus^2\rme^{\rmi p'L}-\oplus^2\rme^{-\rmi p'L}$
(notation in Appendix A). Out of all the terms of the general
homogeneous solution, the choice of $\rme^{\rmi px}$ (for $x>L$) or
$\rme^{-\rmi px}$ (for $x<0$) is dictated by the behaviour of the
solution for $x\rightarrow\pm\infty$ , since a small positive
imaginary part in $p$ is supposed when performing the inverse
Laplace transform leading to $\psi(x,t)\;,\; t>0$ (Appendix D).

However, this expression of the solution is not suited to perform
the inverse Laplace transform back to $\psi(x,t)$ because of the
nontrivial analytical form of the amplitudes above. Instead the GF
approach lets us to express $\bar\psi(x,p)$ as a sum of isolated
pole terms plus other simple terms easier to deal with. To this end,
(2) together with the Green equation for $G(x,x',p)$ can be written
in the form
\begin{eqnarray}
L_x\bar\psi(x,p)-\rmi\alpha \rme^{\rmi kx}&=&0\\
L_xG(x,x',p)-\delta(x-x')&=&0\;,\\\nonumber
\end{eqnarray}
where $G(x,x',p)$ is required to obey RBCs, namely
\begin{equation}
\partial_x G(x,x',p)\!\!\mid_{x=0}\,=-\rmi p\,G(0,x',p)\;\;,\;\;
\partial_x G(x,x',p)\!\!\mid_{x=L}\,=\rmi p\,G(L,x',p)
\end{equation}
It is expected that $G(x,x',p)=L^{-1}_x$ will have poles at
$p=p_n\;$, where the homogeneous equation $L_x\bar\psi=0$ with RBCs
has non-trivial solutions.

The equations (5) and (6) may conveniently be given the short-hand
notation $\Psi\!=\!0$ and $\Gamma\!=\!0$ respectively. Then the
Green method uses the integral equation $\int_{0}^{L}\rmd
x\;[\bar\psi\Gamma-G\Psi]=0$ to obtain $\bar\psi(x,p)$ in terms of
$G(x,x',p)$. Only if $\bar\psi$ and $G$ obey the same
\emph{homogeneous} BCs, the surface terms in this integral cancel
out and one obtains the familiar result
$\bar\psi_{II}(x,p)=\int_{0}^{L}\rmd
x'\,\rmi\alpha\,G(x',x,p)\,\rme^{\rmi kx'}$.

However the BCs for $\bar\psi(x,p)$ (involving $\bar\psi$ and
$\partial_x\bar\psi$ both at $x=0$ and at $x=L$) stemming from (4)
are different from (3) and non-homogeneous. In that case one obtains
a modified expression for the Region $II$, namely
\begin{eqnarray}
\fl\bar\psi_{II}(x,p)\;\;=&&\int_{0}^{L}\rmd x'\,\rmi\alpha\,G(x',x,p)\,\rme^{\rmi kx'}\nonumber\\
&&-\frac{\alpha}{p+k}G(L,x,p)\rme^{\rmi
kL}-\frac{\alpha}{p-k}G(0,x,p)\\\nonumber
\end{eqnarray}
For the external regions $I$ and $III$, the matching of $\bar\psi$
at $x=0$ and $x=L$ determine the coefficients $B$ and $A$
respectively, so that
\begin{eqnarray}
\fl\bar\psi_{I}(x,p)\;\;=&&\rmi\alpha\,\rme^{-\rmi px}\int_{0}^{L}\rmd x'\,\rmi\alpha\,G(x',0,p)\,\rme^{\rmi kx'}\nonumber\\
&&-\frac{\alpha}{p+k}G(L,0,p)\rme^{\rmi kL}\rme^{-\rmi px}-\frac{\alpha}{p-k}G(0,0,p)\rme^{-\rmi px}\nonumber\\
&&-\frac{\rmi\alpha}{p^2-k^2}\rme^{-\rmi
px}+\frac{\rmi\alpha}{p^2-k^2}\rme^{\rmi kx}\\\nonumber
\end{eqnarray}
and
\begin{eqnarray}
\fl\bar\psi_{III}(x,p)\;\;=&&\rmi\alpha\,\rme^{\rmi p(x-L)}\int_{0}^{L}\rmd x'\,\rmi\alpha\,G(x',L,p)\,\rme^{\rmi kx'}\nonumber\\
&&-\frac{\alpha}{p+k}G(L,L,p)\rme^{\rmi p(x-L)}\rme^{\rmi kL}-\frac{\alpha}{p-k}G(0,L,p)\rme^{\rmi p(x-L)}\nonumber\\
&&-\frac{\rmi\alpha}{p^2-k^2}\rme^{\rmi p(x-L)}\rme^{\rmi
kL}+\frac{\rmi\alpha}{p^2-k^2}\rme^{\rmi kx}\\\nonumber
\end{eqnarray}
For comparison, only the last term in the r.h.s. of (8), the last
three ones in (9) and the third one in (10) appear in the shutter
problem.

The crucial advantage of the method is that almost all the Green
functions involved in the equations above can be expanded as a sum
of terms which are simple poles in $p_n$.

\section{Analytical structure of the p-dependent solution}

General theorems \cite{Nussenzveig} and the explicit analytical
derivation (Appendix B) of $G(x,x',p)$ show that $|G(x,x',p)|\sim
1/|p|\,\rightarrow0$ as $|p|\rightarrow\infty$ in the complex plane
for almost any $x\in[0,L]$ and $x'\in[0,L]$ . Then the
Mittag-Leffler theorem tells that
\begin{equation}
G(x,x',p)=\sum_{n}\frac{C_n(x,x')}{p-p_n}
\end{equation}
The exception is for $G(0,0,p)$ and $G(L,L,p)$, the modulus of which
grows as $|p|$ for $|p|\rightarrow\infty$ in the lower half complex
plane $p$ and requires some special care (Appendix C).

The residues in the r.h.s of (11) can be easily computed
\cite{Calderon2} and one finds $C_n(x,x')=u_n(x)u_n(x')/N_n$ , where
the functions $u_n(x)$ belong to the denumerable set of the resonant
solutions satisfying $L_x\,u_n(x)=0$ with RBCs
$$\partial_xu_n(x) \!\!\mid_{x=0}\,=-\rmi p_n\,u_n(0) \;\;,\;\;
\partial_x u_n(x)\!\!\mid_{x=L}\,=\rmi p_n\,u_n(L)$$
and $N_n$ are suitable normalization factors (Appendix A).

The inverse Laplace transform
\begin{equation}
\fl\psi(x,t)=\frac{1}{2\pi \rmi}\int_{c-\rmi \infty}^{c+\rmi
\infty}\rmd s\;\rme^{st}\bar\psi(x,p(s)) = \frac{1}{2\pi
m}\int_{-\infty}^{+\infty}\rmd
p\;p\;\rme^{-\rmi\frac{p^2}{2m}t}\bar\psi(x,p)\;,
\end{equation}
written as an integral over the real momentum variable $p\,$
(Appendix D), leads one to consider the pole expansion of
$p\;\bar\psi(x,p)$ in the integrand, which can be worked out for
each sector $I$, $II$ and $III$. We report here only the result
(Appendix E) for the sectors $II$ and $III$ (sector $I$ is similar
to sector $III$ because of the symmetry of both the potential and
the initial condition):
\begin{eqnarray}
\fl p\;\bar\psi_{II}(x,p)\;\;=&&\rmi\alpha\sum_{n}\frac{p}{p-p_n}\frac{u_n(x)}{N_n}\int_{0}^{L}\rmd x'\,u_n(x')\,\rme^{\rmi kx'}\nonumber\\
&&+\alpha\frac{k}{p+k}G(L,x,-k)\rme^{\rmi kL}-\alpha\,\rme^{\rmi kL}\sum_{n}\frac{p_n}{p-p_n}\frac{1}{k+p_n}\frac{u_n(L)u_n(x)}{N_n}\nonumber\\
&&-\alpha\frac{k}{p-k}G(0,x,k)+\alpha\sum_{n}\frac{p_n}{p-p_n}\frac{1}{k-p_n}\frac{u_n(0)u_n(x)}{N_n}\\\nonumber
\end{eqnarray}

\begin{eqnarray}
\fl p\;\bar\psi_{III}(x,p)\;\;=&&\rmi\alpha\,\rme^{\rmi p(x-L)}\sum_{n}\frac{p}{p-p_n}\frac{u_n(L)}{N_n}\int_{0}^{L}\rmd x'\,u_n(x')\,\rme^{\rmi kx'}\nonumber\\
&&-\alpha\frac{p}{k}G(L,L,0)\rme^{\rmi p(x-L)}\rme^{\rmi kL}+\frac{p^2}{p+k}\frac{\alpha}{k}G(L,L,-k)\rme^{\rmi p(x-L)}\rme^{\rmi kL}\nonumber\\
&&-\alpha\sum_{n}\frac{p^2}{p-p_n}\frac{1}{p_n}\frac{1}{k+p_n}\frac{u_n^2(L)}{N_n}\rme^{\rmi p(x-L)}\rme^{\rmi kL}\nonumber\\
&&+\alpha\sum_{n}\frac{p_n}{p-p_n}\frac{1}{k-p_n}\frac{u_n(0)u_n(L)}{N_n}\rme^{\rmi p(x-L)}+\alpha\frac{k}{k-p}G(0,L,k)\rme^{\rmi p(x-L)}\nonumber\\
&&+\frac{\rmi\alpha}{2}(\frac{1}{p+k}+\frac{1}{p-k})(\rme^{\rmi
kx}-\rme^{\rmi p(x-L)}\rme^{\rmi kL})\\\nonumber
\end{eqnarray}
One recognizes the terms of the shutter problem in the third row of
(13) and in the fourth row of (14).

\section{The time-dependent solution}

For each of the terms above, the integrals stemming from (12) can be
brought to the form of an integral representation of the
(complementary) error function $er\!f\!c(z)$ so that their inverse
Laplace transform can be carried out thoroughly (Appendix F). We
obtain:

\begin{eqnarray}
\fl\psi_{II}(x,t)\;\;=&&-\sum_{n}\frac{u_n(x)}{N_n}(\int_{0}^{L}\rmd x'\,u_n(x')\,\rme^{\rmi kx'})\;[A^{II}_n]\nonumber\\
&&+\rmi kG(L,x,-k)\rme^{\rmi kL}\;[B^{II}]
-\rmi\rme^{\rmi kL}\sum_{n}\frac{p_n}{k+p_n}\frac{u_n(L)u_n(x)}{N_n}\;[B^{II}_n]\nonumber\\
&&-\rmi kG(0,x,k)\;[S^{II}] +\rmi
\sum_{n}\frac{p_n}{k-p_n}\frac{u_n(0)u_n(x)}{N_n}\;[S^{II}_n]\\\nonumber
\end{eqnarray}
where the factors in square brackets embody the time
($\tau=t\,/\,2m$) dependence of the solution and are the result of
the integrations over the momentum $p$. Explicitly
\begin{eqnarray}
A^{II}_n&=&-p_n\rme^{-\rmi\tau{p_n^2}}\;er\!f\!c(\rmi\sqrt{\rmi\tau})-\frac{\rme^{-\rmi\frac{\pi}{4}}}{\sqrt{\pi\tau}}\nonumber\\
B^{II}&=&-\rme^{-\rmi\tau{k^2}}\;er\!f\!c(-\rmi\sqrt{\rmi\tau}k)\nonumber\\
B^{II}_n&=&-\rme^{-\rmi\tau{p_n^2}}\;er\!f\!c(\rmi\sqrt{\rmi\tau}p_n)\nonumber\\
S^{II}&=&-\rme^{-\rmi\tau{k^2}}\;er\!f\!c(\rmi\sqrt{\rmi\tau}k)\nonumber\\
S^{II}_n&=&-\rme^{-\rmi\tau{p_n^2}}\;er\!f\!c(\rmi\sqrt{\rmi\tau}p_n)\\\nonumber
\end{eqnarray}
Likewise
\begin{eqnarray}
\fl\psi_{III}(x,t)\;\;=&&-\sum_{n}\frac{u_n(L)}{N_n}(\int_{0}^{L}\rmd x'\,u_n(x')\,\rme^{\rmi kx'})\;[A^{III}_n]\nonumber\\
&&-iG(L,L,0)\frac{\rme^{\rmi kL}}{k}\;[B^{III}_0]
+iG(L,L,-k)\frac{\rme^{\rmi kL}}{k}\;[B^{III}_{-k}]\nonumber\\
&&-\rmi\rme^{\rmi kL}\sum_{n}\frac{1}{p_n}\frac{1}{k+p_n}\frac{u_n^2(L)}{N_n}\;[B^{III}_n]\nonumber\\
&&-\rmi kG(0,L,k)\;[S^{III}]+\rmi
\sum_{n}\frac{p_n}{k-p_n}\frac{u_n(0)u_n(L)}{N_n}\;[S^{III}_n]
\hskip 3cm\nonumber\\
&&-\frac{1}{2}\rme^{\rmi kx}\;[C^{III}_1] -\frac{1}{2}\rme^{\rmi
kx}\;[C^{III}_2] +\frac{1}{2}\rme^{\rmi
kL}\;[C^{III}_3]+\frac{1}{2}\rme^{\rmi kL}\;[C^{III}_4]\\\nonumber
\end{eqnarray}
where
\begin{eqnarray}
A^{III}_n&=&-\rme^{\rmi\frac{(x-L)^2}{4\tau}}(p_n\rme^{y_n^2}\;er\!f\!c(y_n)+\frac{\rme^{\rmi\frac{\pi}{4}}}{\sqrt{\pi\tau}})\nonumber\\
B^{III}_0&=&-\frac{\rme^{\rmi\frac{\pi}{4}}}{\sqrt{\pi\tau}}\frac{x-L}{2\tau}\rme^{\rmi\frac{(x-L)^2}{4\tau}}\nonumber\\
B^{III}_{-k}&=&-\rme^{\rmi\frac{(x-L)^2}{4\tau}}(k^2\rme^{y_k^2}\;er\!f\!c(y_{-k})+\frac{\rme^{\rmi\frac{\pi}{4}}}{\sqrt{\pi\tau}}(-k+\frac{x-L}{2\tau}))\nonumber\\
B^{III}_n&=&-\rme^{\rmi\frac{(x-L)^2}{4\tau}}(p_n^2\rme^{y_n^2}\;er\!f\!c(y_n)+
\frac{\rme^{\rmi\frac{\pi}{4}}}{\sqrt{\pi\tau}}(p_n+\frac{x-L}{2\tau}))\nonumber\\
S^{III}&=&-\rme^{\rmi\frac{(x-L)^2}{4\tau}}\rme^{y_k^2}\;er\!f\!c(y_k)\nonumber\\
S^{III}_n&=&-\rme^{\rmi\frac{(x-L)^2}{4\tau}}\rme^{y_n^2}\;er\!f\!c(y_n)\nonumber\\
C^{III}_1&=&-\rme^{-\rmi\tau{k^2}}\;er\!f\!c(-\rmi\sqrt{\rmi\tau}k)\nonumber\\
C^{III}_2&=&-\rme^{-\rmi\tau{k^2}}\;er\!f\!c(\rmi\sqrt{\rmi\tau}k)\nonumber\\
C^{III}_3&=&-\rme^{\rmi\frac{(x-L)^2}{4\tau}}\rme^{y_{-k}^2}\;er\!f\!c(y_{-k})\nonumber\\
C^{III}_4&=&-\rme^{\rmi\frac{(x-L)^2}{4\tau}}\rme^{y_k^2}\;er\!f\!c(y_k)\\\nonumber
\end{eqnarray}
We have introduced the variables
$y_q=\rme^{-\rmi\frac{\pi}{4}}(4\tau)^{-\frac{1}{2}}(x-L-2\tau q)$
for $q=p_n,k,-k$. In both equations the terms $[S]$ are the ones
arising in the shutter problem. The particular values of the Green
function involved in eqs.(15) and (17) are calculated in Appendix B.

\section{The short time limit}
The $t\rightarrow0$ limit is interesting both as a check of the
calculation above and for the study of the scattered wave at short
times. Here we aim only to recover the initial wave function
$\psi(x,0)= \rme^{\rmi kx}$ at $t=0$, which must happen in each of
the sectors $I$, $II$ and $III$. The sector $I$ similar to $III$, so
we refrain from calculating it.

Careful use of the limiting values of $er\!f\!c(z)$ and/or
$w(z)=\rme^{-z^2}er\!f\!c(-\rmi z)$ for $z\rightarrow0$ and for
$z\rightarrow\infty$ in different directions of the complex $z$
plane must be made. Notice that $y_q$ tends to $\infty$ in different
directions for different $q$. Also the properties of the set of
resonant functions $u_n(x)$ as a basis of the space of solutions are
crucial (Appendix G).

\subsection{Region II}
For $\tau\rightarrow0$ the factors $[B^{II}]$, $[B^{II}_n]$,
$[S^{II}]$ and $[S^{II}_n]$ tend to the value $-1$, so that the
$[B]$ terms in (15) yield
\begin{eqnarray}
\fl\rmi kG(L,x,-k)\rme^{\rmi kL}-&&\rmi\rme^{\rmi kL}\sum_{n}\frac{p_n}{k+p_n}\frac{u_n(L)u_n(x)}{N_n}\nonumber\\
&&=\rmi k\rme^{\rmi kL}\sum_{n}\frac{1}{-k-p_n}\frac{u_n(L)u_n(x)}{N_n}-\rmi\rme^{\rmi kL}\sum_{n}\frac{p_n}{k+p_n}\frac{u_n(L)u_n(x)}{N_n}\nonumber\\
&&=-\rmi\rme^{\rmi kL}\sum_{n}\frac{u_n(L)u_n(x)}{N_n}=0\\\nonumber
\end{eqnarray}
and likewise for the $[S]$ terms:
\begin{eqnarray}
\fl\rmi kG(0,x,k)-\rmi \sum_{n}&&\frac{p_n}{k-p_n}\frac{u_n(0)u_n(x)}{N_n}\nonumber\\
&&=\rmi k\sum_{n}\frac{1}{k-p_n}\frac{u_n(0)u_n(x)}{N_n}-\rmi \sum_{n}\frac{p_n}{k-p_n}\frac{u_n(0)u_n(x)}{N_n}\nonumber\\
&&=\rmi \sum_{n}\frac{u_n(0)u_n(x)}{N_n}=0\\\nonumber
\end{eqnarray}
The limit of the factor $[A^{II}_n]$ is not readable in (16) since
it seemingly blows up for $\tau\rightarrow0$. Actually, the changes
of variables done to perform the integration leading to this form of
$[A^{II}_n]$ become singular in this limit. The right result can be
reached by the following reasoning: For $\tau\rightarrow0$, an
increasingly large range of values of $p$ (as compared to $Re\,p_n$)
does contribute to the integral $\int_{-\infty}^{+\infty}\rmd
p\;\rme^{-\rmi\tau p^2}p/(p-p_n)$. Therefore its limiting value
$(\pi\,/\,\rmi\tau)^{\frac{1}{2}}$ does not depend on $p_n$ and
becomes a common factor outside the first sum in (15). Then one
recovers the sum $\sum u_n(x)u_n(x')/N_n= \delta(x-x')$, as expected
from the $t\rightarrow0$ limit of the Green function $G(x,x',t)$.

Therefore, for $0\leq x\leq L\;$,
$$lim_{t\rightarrow0}\;\psi_{II}(x,t)= \rme^{\rmi kx}\equiv\psi_o(x)$$
as required.

Asymptotic expressions for $er\!f\!c(z)$ and $w(z)$ can be used to
obtain approximations to the form of $\psi_{II}(x,t)$ for small
values of $t$, a task that will be faced elsewhere.

\subsection{Region III}

The limit must be studied term by term. The integral
$\int_{-\infty}^{+\infty}\rmd p\;\rme^{-\rmi\tau p^2}\rme^{\rmi
p(x-L)}p/(p-p_n)$ has now an extra exponential factor which has a
regularizing effect. As a result, the limit can be taken directly in
$[A^{III}_n]$ as given in (18): the term $\rme^{y_n^2}er\!f\!c(y_n)$
vanishes and
$(\pi\tau)^{-\frac{1}{2}}\rme^{\rmi\frac{(x-L)^2}{4\tau}}$
approaches a distribution concentrated in $x=L$, namely
$\delta(x-L)$. Thus it vanishes for $x>L$.

An analogous reasoning shows that $[B^{III}_0]$, $[B^{III}_{-k}]$
and $[B^{III}_n]$ yield derivatives of $\delta(x-L)$ and also vanish
for $x>L$.

The factors $[S^{III}]$, $[S^{III}_n]$, $[C^{III}_3]$ and
$[C^{III}_4]$ vanish exactly, whereas $[C^{III}_1]$ and
$[C^{III}_2]$ tend to $-1$. From (17) we then see that the final
result is that also
$$lim_{t\rightarrow0}\;\psi_{III}(x,t)= \rme^{\rmi kx}\equiv\psi_o(x)$$
for $x>L$, as required.

The solution $\psi_{I}(x,t)$ at the left of the barrier will have a
structure similar to (17) and the same limit above.

\section{The large time limit}

For $\tau\rightarrow\infty$ we see that also $y_q\rightarrow\infty$
as before, but in still different directions of the complex plane
for the different $q$.

In the internal Region II, the factors $[A^{II}_n]$, $[B^{II}]$,
$[B^{II}_n]$ and $[S^{II}_n]$ vanish, whereas $[S^{II}]\rightarrow
-2\rme^{-\rmi\tau k^2}$. Therefore, for $0\leq x\leq L$,
\begin{eqnarray}
lim_{t\rightarrow\infty}\;\psi_{II}(x,t)&=&2\rmi k\;G(0,x,k)\rme^{-\rmi\tau k^2}\nonumber\\
&=&\phi^{in}_r(x)\rme^{-\rmi E_kt}\\\nonumber
\end{eqnarray}
that is the stationary scattering {\it in} solution, where
$E_k=k^2/\,2m\;$ (see Appendices B and H). Thus the infinite plane
wave evolves into the same final state of the shutter initial
condition.

For the external Region III, the only non-vanishing factors are
\begin{eqnarray}
S^{III}&\rightarrow&-2\;\rme^{\rmi k(x-L)}\rme^{-\rmi\tau k^2}\nonumber\\
C^{III}_2&\rightarrow&-2\;\rme^{-\rmi\tau k^2}\nonumber\\
C^{III}_4&\rightarrow&-2\;\rme^{\rmi k(x-L)}\rme^{-\rmi\tau k^2}\nonumber\\
\end{eqnarray}
Then the terms corresponding to $C^{III}_2$ and $C^{III}_4$ in (17)
cancel each other and, again, one is left with the same asymptotic
solution of the shutter
\begin{eqnarray}
lim_{t\rightarrow\infty}\;\psi_{III}(x,t)&=&2\rmi k\;G(0,L,k)\rme^{\rmi k(x-L)}\rme^{-\rmi\tau k^2}\nonumber\\
&=&T(k)\rme^{\rmi kx}\rme^{-\rmi E_kt}\nonumber\\
&=&\phi^{in}_r(x)\rme^{-\rmi E_kt}\\\nonumber
\end{eqnarray}

The finite time behaviour is thus made up of transient modes which
quickly disappear leaving (for $k>0$) the scattering asymptotic
solution with outgoing BC at $x=L$.

\section{Conclusions}

A solution for the time evolution of an infinite plane wave in the
presence of a simple square barrier has been worked out for each of
the regions of the potential and Section 4 is the main result of
this paper. Among other terms, this solution contains a sum of
explicit analytical contributions corresponding to each of the
(infinitely many) resonance poles. As in previous works in related
problems \cite{Muga1} \cite{Calderon1}, only the location of these
poles needs to be obtained by numerical methods.

In the solution obtained the shutter terms have been pinpointed.
Similarly, the contributions to the evolved wave function coming
from the segments of the initial wave function lying inside and at
the right of the barrier can be identified.

As the main application of this knowledge we envisage the
possibility of studying the enhancement or the suppression of the
transmission of the single Fourier components of any realistic wave
packet. This should provide new detailed insight on interesting
phenomena like the super-luminal tunnelling \cite{Leon}, the
breakdown of energy conservation \cite{Muga2} by transient
interference in wave packet collisions with barriers \cite{Muga3} or
the rising of forerunners.

Other side results of this work are more mathematical and
preliminary for further study. The exact analytic expression for the
GF of the square barrier subject to RBCs has been worked out. Some
particular values of it are involved in the time-dependent solution
obtained above, but its knowledge is also useful for computing in
each region the extra terms in general resonance pole expansions of
wave functions in order to implement, with an explicit example, the
related question of the completeness of the set of the resonant
solutions \cite{Calderon3}. We have relied on the adoption of
different boundary conditions for the GF and for the solutions of
the (Laplace transformed) evolution equations. The link between this
point and the properties of completeness of the solutions will be
addressed elsewhere.

On more physical grounds, the explicit solution obtained for finite
time is useful for deriving approximations valid for short times,
hence for the study of transient structures and forerunners. An
immediate result of the work is that the large time limit yields the
same stationary solution of the shutter. This again shows that the
resonances contribute only to transient structures of the scattered
wave.

\ack{Work supported by MEC projects BFM2002-00834 and FIS2005-05304.
The authors are indebted to J. Le\'on for suggestions and helpful
discussions. J. Julve acknowledges the hospitality of the
Dipartimento di Fisica dell'Universit\`a di Bologna, Italy, where
part of this work was done.}

\appendix

\section{Resonant solutions}
We adopt units such that $\hbar=1$ and define the
 differential operator
$$L_x\equiv [\frac{\partial^2}{\partial x^2}+p^2-2m\,V(x)]\,.$$ A
piece-wise general solution for the homogeneous SOLDE
$\;L_x\,u(x)=0$ is
$$u(x,p)=\theta(-x)B\,\rme^{-\rmi px}+\theta(x)\theta(L-x)(M\,\rme^{\rmi p'x}+N\,\rme^{-\rmi p'x})
+\theta(x-L)\,A\,\rme^{\rmi px}\;,$$ which has already built-in the
RBCs (3). Choosing $B(p)$ arbitrary, the matching conditions at
$x=0$ and at $x=L$ yield the amplitudes
\begin{eqnarray}
M(p)&=&-\frac{\ominus}{2p'}B(p)\nonumber\\
N(p)&=&-\frac{\oplus}{2p'}B(p)\nonumber\\
A(p)&=&-\frac{1}{2p'}(\oplus \rme^{-\rmi\oplus L}-\ominus
\rme^{-\rmi\ominus L})B(p)\;,\\\nonumber
\end{eqnarray}
where $p'\equiv\sqrt{p^2-2mV}\;$, $\oplus\equiv p+p'\;$ and
$\ominus\equiv p-p'$ , plus an extra condition on $p$ which is
proportional to
\begin{equation}
\oplus^2\rme^{-\rmi\oplus L}-\ominus^2\rme^{-\rmi\ominus L}=0
\end{equation}
This transcendent equation has a denumerable set of solutions
$p_n\;$ lying in the lower complex plane $p\,$. The common
denominator $D(p)\equiv\ominus^2\rme^{\rmi p'L}-\oplus^2\rme^{-\rmi
p'L}$ in the amplitudes of (4) is just proportional to the l.h.s. of
(A.2). One can check that if $p_n$ is a solution, then $-p_n^*$ is
too, so that these values are in symmetrical locations with respect
to the imaginary axis. We let the label $n$ take integer values
($n\neq0$), with the growing positive $n$ indicating the $p_n$ with
growing real positive part and $p_{-n}\equiv -p_n^*$. One finds that
$-\frac{\pi}{4}<ar\!g\,p_n<0$ and
$\pi<ar\!g\,p_{-n}<\frac{5\pi}{4}$.

The simplest writing of the solution (24) is obtained for
$B(p)=-2p'$, namely
\begin{eqnarray}
u_n(x)=&&\theta(-x)(-2p'_n)\rme^{-\rmi p_nx}\nonumber\\
&+&\theta(x)\theta(L-x)(\ominus_n\,\rme^{\rmi p'_nx}-\oplus_n\,\rme^{-\rmi p'_nx})\nonumber\\
&+&\theta(x-L)\,(\ominus_n\,\rme^{-\rmi\ominus_nL}-\oplus_n\,\rme^{-\rmi\oplus_nL})\,\rme^{\rmi
p_nx}\;,\\\nonumber
\end{eqnarray}

Notice that the solution (4) should yield the outgoing resonant
ones, which exist only for the momenta $p_n$ , when we let the terms
involving the initial condition $\rme^{\rmi kx}$ (namely the
particular solution of the inhomogeneous equation) vanish. In fact,
one sees that for $p\rightarrow p_n$ , these terms belong to the
regular part of (4), with corresponding inhomogeneous BCs, while the
pole part yields the resonant solution.

The crucial role of the BCs in solving the $\rm Schr\ddot{o}dinger$
equation \cite{Madrid2} is behind the fact that different BCs like
the RBCs, obeyed by the resonant solutions and by the Gf (causing
its resonance poles), on one side, and the ones characterizing the
scattering solutions (whose amplitudes display the S-matrix poles)
on the other, lead to the same poles.

The residues $C_n(x,x')=u_n(x)u_n(x')/N_n$ obtained in (11)
correspond to the following choice of (complex) "norm"
\cite{Calderon2} \begin{equation}
N_n=\rmi(u^2_n(0)+u^2_n(L))+2p_n\int^L_0\rmd x\;u^2_n(x)
\end{equation}
which takes the value $N_n=-8mV(p_n\,L+2\rmi)$ for the solutions
(A.3). This "norm" is somehow related to the usual one in Hilbert
space: when applied to ordinary square-integrable wave functions
$\psi(x)\in {\cal L}^2$, replacing the squares by the square modulus
and letting $0$ and $L$ recede respectively to $-\infty$ and to
$+\infty$, equation (A.4) approaches the usual norm (up to a
factor). For the resonant solutions however, $N_n$ diverges in this
limit .

\section{Analytical Green function}

The analytical solution to the Green equation
$\;L_xG(x,y,p)=\delta(x-y)\;$ for $x$ and $y$ in the interval
$[0,L]\,$, and obeying the RBCs (7), can be directly constructed:

\begin{eqnarray}
\fl G(x,y,p)=\frac{\rmi}{2p'}\frac{1}{D(p)}\;&\{&\;2mV\,(\rme^{\rmi p'(L-(x+y))}+\rme^{-\rmi p'(L-(x+y))})\\
&&\!\!\!\!-\oplus^2\rme^{-\rmi p'L}(\theta(y-x)\rme^{\rmi p'(y-x)}+\theta(x-y)\rme^{\rmi p'(x-y)})\nonumber\\
&&\!\!\!\!-\ominus^2\rme^{\rmi p'L}(\theta(y-x)\rme^{-\rmi
p'(y-x)}+\theta(x-y)\rme^{-\rmi p'(x-y)})\,\}\;,\nonumber\\\nonumber
\end{eqnarray}
where the symmetry $x\leftrightarrow y$ is explicit and the zeroes
of $D(p)$ give the expected simple poles.

Particular useful values of (B.1) are:

\begin{eqnarray}
&&G(0,x,p)=\frac{\rmi}{D(p)}(\ominus \rme^{\rmi p'(L-x)}-\oplus \rme^{-\rmi p'(L-x)})\nonumber\\
&&G(L,x,p)=\frac{\rmi}{D(p)}(\ominus \rme^{\rmi p'x}-\oplus \rme^{-\rmi p'x}) \nonumber\\
&&G(0,L,p)=-\rmi\frac{2p'}{D(p)} \nonumber\\
&&G(0,0,p)=G(L,L,p)= \frac{\rmi}{D(p)}(\ominus \rme^{\rmi p'L}-\oplus \rme^{-\rmi p'L})\hskip 3cm\nonumber\\
&&G(x,y,0)=\frac{-1}{2\sqrt{2mV}Sinh[L\sqrt{2mV}]}\;\{Cosh[\sqrt{2mV}(L-(x+y))]\nonumber\\
&&\hskip 4.5cm+\theta(y-x)Cosh[\sqrt{2mV}(L-(y-x))]\nonumber\\
&&\hskip 4.5cm+\theta(x-y)Cosh[\sqrt{2mV}(L-(x-y))]\}\nonumber\\
&&G(0,0,0)=G(L,L,0)=-\frac{1}{\sqrt{2mV}}\;Coth[L\sqrt{2mV}] \nonumber\\
&&\partial_pG(0,0,p)\!\mid_{p=0}=-\frac{\rmi}{4mV}\,\frac{3+Cosh[2L\sqrt{2mV}]}{Sinh[L\sqrt{2mV}]}
\\\nonumber
\end{eqnarray}

The limit $|p|\rightarrow\infty$ can be directly read out in the
above expressions. It can be seen that in all the cases the Green
function vanishes as $\;1/p\;$ or faster in any direction of the
complex plane, with the only exception of $G(0,0,p)$ and
$G(L,L,p)\,$, which grow as $p$ in the lower half plane, though they
still decrease as $\;1/p\;$ in the real axis.

The particular cases $G(0,x,p)$ and $G(L,x,p)$ are related to the
scattering solutions $\phi^{in}(x)$ (see Appendix H). From the
integral formula $\;\int^L_0\rmd
x\,[\phi\,(L_xG-\delta(x-x'))-G\,L_x\phi]=0\;$ and the use of the
BCs for $\phi$ and the RBCs for $G\,$, one obtains \cite{Calderon4}
\begin{eqnarray}
G(0,x,p)&=&\frac{-\rmi}{2p}\;\sqrt{2\pi}\sqrt{\frac{p}{m}}\;\phi^{in}_r(x)\hskip 1cm (0<x\leq L)\\
G(L,x,p)&=&\frac{-\rmi}{2p}\;\sqrt{2\pi}\sqrt{\frac{p}{m}}\;\phi^{in}_l(x)\hskip
1cm (0\leq x<L)\\\nonumber
\end{eqnarray}

\section{Pole expansion and substractions}
Whenever the Green function $G(x,x',p)$ vanishes as $\;1/p\;$ in the
limit $|p|\rightarrow\infty$, the Mittag-Leffler theorem applies and
the validity of the pole expansion (11) is assured. This is not the
case for $G(0,0,p)$ and $G(L,L,p)\,$, which grow as $p$ in the lower
half plane, and a substraction technique must be used
\cite{Nussenzveig}. This is performed by using the contour integral

\begin{equation}
0=\frac{1}{2\pi \rmi}\oint_{\Gamma}\rmd
z\,\frac{1}{z^2}\frac{G(x,x',z)}{z-p}
\end{equation}
where the contour $\Gamma\equiv\{C_o,C_p\,,C_n,C_s\}$ consists of
small closed paths counterclock-wise encircling the poles of the
integrand (namely $C_o$ around the double pole at $z=0$, $C_p$
around $p$ and $C_n$ around each $p_n$) plus a large circle $C_s$
clock-wise encircling all these poles. The factor $z^{-2}$ has been
introduced to assure that the integrand goes as $z^{-2}$ even for
$G(0,0,p)$ and $G(L,L,p)\,$ so that the integral over $C_s$ vanishes
when the radius of the circle grows to $\infty$.

From (C.1) one readily obtains (see also \cite{Calderon5})
\begin{equation}
\fl
G(x,x',p)=p^2\sum^{\infty}_n\frac{1}{N_n}\frac{1}{p^2_n}\frac{u_n(x)u_n(x')}{p-p_n}
+[G(x,x',0)+p\;\partial_pG(x,x',p)\!\mid_{p=0}]
\end{equation}
When the pole expansion (11) holds, (C.2) becomes a trivial
identity, whereas in the case of $G(0,0,p)$ and $G(L,L,p)\,$ a
holomorphic (linearly growing with $|p|$ ) part remains.

\section{Inverse Laplace transform}

The inverse Laplace transform in the variable $s$ in (12), involves
an integration in the complex $s$-plane along a line parallel to the
imaginary axis with $Re\,c>0$. In the $p$-plane (recall
$p=\sqrt{\rmi}\sqrt{2m\,s}$), this path translates to a
hyperbola-like one with asymptotes in the positive real and
imaginary axis. For $t>0\,$, the factor
$\rme^{-\rmi\frac{p^2}{2m}t}$ assures that it can be deformed into
an integration from $+\infty$ to $-\infty$ along (and slightly
above) the real axis if the integrand has poles only in the lower
half plane (and on the real axis). For $t<0\,$, the path can be
closed along a quarter circle of large radius in the 1st quadrant,
thus enclosing a region without poles and giving $\psi(x,t)=0$,
consistently with the Laplace method and causality
\cite{Nussenzveig}.

\section{Pole expansion of $p\;\bar\psi(x,p)$}
We rewrite (8) as
\begin{equation}
\fl\bar\psi_{II}(x,p)\;=\;\rmi\alpha\int_{0}^{L}\rmd
x'\,f^{II}_1(x',x,p)\,\rme^{\rmi kx'} -\alpha\;\rme^{\rmi
kL}f^{II}_2(x,p)-\alpha\, f^{II}_3(x,p)\\\nonumber
\end{equation}
where
$$\;f^{II}_1(x',x,p)=G(x',x,p)\;,
\;f^{II}_2(x,p)=\frac{G(L,x,p)}{p+k}\; {\rm and}
\;f^{II}_3(x,p)=\frac{G(0,x,p)}{p-k}\;.$$ According to the
asymptotic behaviour of the $\;f^{II}_i(p)\;$ ($i=1,2,3$) above we
consider the contour integrals
\begin{equation}
0=\frac{1}{2\pi \rmi}\oint_{\Gamma_i}\rmd
z\,\frac{z^{n_i}}{z-p}f^{II}_i(z)\nonumber
\end{equation}
where the values $n_1=0$, and $n_2=n_3=1\;$ are assigned so that the
integrand of (E.2) behaves as $z^{-2}$ for large $z\;$. The contours
$\Gamma_i$ include circles around the poles in the respective
integrand (namely $p\,$, $p_n$, $k$ or $-k\;$) plus the large circle
$C_s\,$, as in Appendix C. For $i=1$, equation (E.2) yields
$f^{II}_1(p)$, which must be multiplied by $p$ later on, whereas for
$i=2,3$ one directly obtains $p\,f^{II}_i(p)$. Collecting these
results and using the pole expansion of the Green function, (13) is
obtained.

The derivation of (14) follows the same lines. Here we define
$$\;f^{III}_1(p)=G(x',L,p)\;,
\;f^{III}_2(p)=\frac{G(L,L,p)}{p+k}\; {\rm and}
\;f^{III}_3(p)=\frac{G(0,L,p)}{p-k}\;,$$ and the powers required are
$n_1=0$, $n_2=-1$ and $n_3=1\;$. Notice that the negative power
$n_2=-1$, needed by the asymptotic behaviour of $G(L,L,z)\,$,
introduces an extra pole at $z=0$ and is related to the discussion
of App.C.

\section{{\it erfc(z)}-related integrals}
In the computation of $\psi_{II}(x,t)$ and $\psi_{III}(x,t)$, the
following types of integrals arise:

\begin{eqnarray}
I(x)&=&\frac{1}{\rmi\pi}\int^{+\infty}_{-\infty}\rmd p\;p\;\rme^{-\rmi\tau p^2}\rme^{\rmi(x-L)p}\\
I_0(x,q)&=&\frac{1}{\rmi\pi}\int^{+\infty}_{-\infty}\rmd p\;\frac{1}{p-q}\,\rme^{-\rmi\tau p^2}\rme^{\rmi(x-L)p}\equiv -2\,M(x-L,q,t)\\
I_1(x,q)&=&\frac{1}{\rmi\pi}\int^{+\infty}_{-\infty}\rmd p\;\frac{p}{p-q}\,\rme^{-\rmi\tau p^2}\rme^{\rmi(x-L)p}=-\rmi\frac{\partial}{\partial x}I_0(x,q)\\
I_2(x,q)&=&\frac{1}{\rmi\pi}\int^{+\infty}_{-\infty}\rmd
p\;\frac{p^2}{p-q}\,\rme^{-\rmi\tau
p^2}\rme^{\rmi(x-L)p}=(-\rmi\frac{\partial}{\partial
x})^2I_0(x,q)\;,\\\nonumber
\end{eqnarray}
where

$$M(x,q,t)\equiv\frac{-1}{2\pi
\rmi}\int^{+\infty}_{-\infty}\rmd
p\;\frac{1}{p-q}\,\rme^{-\rmi\frac{t}{2m} p^2}\rme^{ixp}$$ is the
Moshinsky function, often found in the literature.

They can be related to the integral representation of the $error\,
f\!unction$ \cite{Stegun} \begin{equation} w(z)\equiv \rme^{-z^2}
er\!f\!c(-\rmi z)=
-\frac{\rmi}{\pi}\int^{+\infty}_{-\infty}du\;\frac{\rme^{-u^2}}{u-z}
\hskip 2cm Im\,z>0
\end{equation}
so that
$$M(x,q,t)=\frac{1}{2}\rme^{\rmi\frac{m}{2t}x^2}w(\rmi\rme^{-\rmi\frac{\pi}{4}}\sqrt{\frac{m}{2t}}(x-\frac{t}{m}
q))\;.$$

When converting the exponent $\rmi\tau p^2$ into $u^2$, namely
$u=\sqrt{\rmi\tau}p\,$, an integration along a line running between
$\pm \sqrt{\rmi}\,\infty$ results. In the case $q=p_n\,$, rotating
this path by an angle $\pi/4$ towards the real axis crosses the pole
$\sqrt{\rmi\tau}\,p_n$, contributing a residue. For $q=\pm k$, a
small negative imaginary part is supposed (see Appendix D). The case
$Im\,z<0$ can be tackled by using the property $w(z^*)=w^*(-z)$.

The results of the integrals above, as given in Section 4, are

\begin{eqnarray}
A^{II}_n&=&I_1(L,p_n)\nonumber\\
A^{III}_n&=&I_1(x,p_n)\nonumber\\
B^{III}_0&=&I(x)\nonumber\\
B^{III}_{-k}&=&I_2(x,-k)\nonumber\\
B^{III}_n&=&I_2(x,p_n)\nonumber\\
S^{III}&=&I_0(x,k)\nonumber\\
S^{III}_n&=&I_0(x,p_n)\nonumber\\
B^{II}=C^{III}_1&=&I_0(L,-k)\nonumber\\
S^{II}=C^{III}_2&=&I_0(L,k)\nonumber\\
C^{III}_3&=&I_0(x,-k)\nonumber\\
C^{III}_4&=&I_0(x,k)\nonumber\\
B^{II}_n=S^{II}_n&=&I_0(L,p_n)\\\nonumber
\end{eqnarray}

\section{$t\rightarrow 0,\infty$ limits of $\rme^{y^2_q}\,er\!f\!c(y_q)$}

Being
$\;y_q\equiv\frac{\rme^{-\rmi\frac{\pi}{4}}}{\sqrt{4\tau}}(x-L-2\tau
q)\,$, where $\tau\equiv\frac{t}{2m}$ , we see that, for
$t\rightarrow 0\,$,
$$y_q\sim \rme^{-\rmi\frac{\pi}{4}}\sqrt{\frac{m}{2t}}(x-L)\rightarrow \rme^{-\rmi\frac{\pi}{4}}\cdot\infty\;,$$
regardless of $q$ being equal to $\pm k$, $p_n\,$ or $p_{-n}\,$.

For $t\rightarrow \infty\,$ one has
$$y_q\sim \rme^{\rmi\,\frac{3\pi}{4}}\sqrt{\frac{t}{2m}}\;\;q\;\rightarrow\; \rme^{\rmi\phi_q}\cdot\infty\;,$$
where $\;\frac{\pi}{2}<\phi_n<\frac{3\pi}{4}\;\;$,
$\;-\frac{\pi}{4}<\phi_{-n}<0\;\;$,  $\phi_k=\frac{3\pi}{4}\;\;$ and
$\;\phi_{-k}=-\frac{\pi}{4}\;\;$.

The following asymptotic formulas \cite{Stegun} apply:
$$lim_{z\rightarrow\infty}\;er\!f\!c(z)=0\hskip 2cm|ar\!g\,z|<\frac{\pi}{4}$$
$$lim_{z\rightarrow\infty}\;\rme^{z^2}er\!f\!c(z)\sim\frac{1}{\sqrt{\pi}}\frac{1}{z}(1+\sum_{m=1}^{\infty}\frac{C_m}{z^{2m}})\rightarrow 0\hskip 2cm|ar\!g\,z|<\frac{3\pi}{4}$$
Only the case $\phi_k=\frac{3\pi}{4}\;$ is out of their range, in
which case the relationship
$$\rme^{y^2_k}\,er\!f\!c(y_k)=2\rme^{y^2_k}-\rme^{-y^2_k}\,er\!f\!c(-y_k)\;,$$
where $|ar\!g(-y_k)|=\frac{\pi}{4}\,$, is helpful.

\section{Scattering solutions}

The usual scattering solutions of $L_x\bar\psi(x,p)=0$ for
continuous real energies $E=p^2/\,2m\,>0\;$ may be characterized
according to the homogeneous BC they obey. Contrary to the resonant
solutions, which are subject to {\it two} BCs,  each one of the
scattering solutions solution is subject to {\it one} homogeneous
BC, alternatively at $x=0$ or at $x=L$ :
\begin{eqnarray}
\partial_x\phi^{in}_r(x)\!|_L&=&\rmi p\,\phi^{in}_r(L)\nonumber\\
\partial_x\phi^{in}_l(x)\!|_0&=&-\rmi p\,\phi^{in}_l(0)\nonumber\\
\partial_x\phi^{out}_r(x)\!|_0&=&\rmi p\,\phi^{in}_r(0)\nonumber\\
\partial_x\phi^{out}_l(x)\!|_L&=&-\rmi p\,\phi^{in}_l(L)\;,\\\nonumber
\end{eqnarray}
where $r$($l$) labels the right(left)-moving character of the
impinging wave. Notice that the reverse notation is often found in
the literature, with  $r$($l$) indicating the wave coming from the
right(left) of the barrier.

These solutions, for instance
$$\phi^{in}_r(x,p)=\frac{1}{\sqrt{2\pi}}\sqrt{\frac{m}{p}}\;[\;\theta(-x)(\rme^{\rmi px}+R\,\rme^{-\rmi px})\nonumber\\
+\theta(x)\theta(L-x)(P\,\rme^{\rmi p'x}+Q\,\rme^{-\rmi p'x})
+\theta(x-L)\,T\,\rme^{\rmi px}\,]\,,$$  are $\delta$-normalized in
energy.

\end{document}